\documentclass[a4paper]{elsarticle}

\usepackage{moreverb}
\usepackage[colorlinks,bookmarksopen,bookmarksnumbered,citecolor=red,urlcolor=red]{hyperref}

\usepackage{url}
\usepackage{xfrac}
\usepackage{a4wide} 
\usepackage{graphicx}
\usepackage{listings}
\usepackage[utf8]{inputenc}
\usepackage{color}
\usepackage{booktabs}
\usepackage{paralist}
\usepackage{inconsolata}
\usepackage{floatrow}
\usepackage{algpseudocode}
\usepackage[]{amsmath}
\usepackage[]{amssymb}

\usepackage[norelsize]{algorithm2e}

\graphicspath{{figures/}}

\newfloatcommand{capbtabbox}{table}[][\FBwidth]

\definecolor{light-gray}{gray}{0.85}
\definecolor{dkgreen}{rgb}{0,0.35,0}

\lstdefinelanguage{Skeletons}{
	keywords={
  		KernelWrapper,Stream,Pipeline,MapReduce,split,merge,Loop,BufferData,FinalData,
  		SingletonData,write,wait,IWorkData,IExecutable,IFuture,Vector, Map, pipeline, kernel, OpenCLKernel
	},
	morecomment = [l]{//},
	morecomment = [l]{///},
	morecomment = [s]{/*}{*/},
    commentstyle=\color{dkgreen}      % comment style
}

\begin{document}

 \begin{frontmatter}

\title{Execution of Compound Multi-Kernel OpenCL Computations in Multi-CPU/Multi-GPU Environments}

\tnotetext[label0]{Preprint of the article published in a special issue of John Wiley \& Sons' Concurrency and Computation: Practice and Experience}

\tnotetext[label1]{This work was partially funded by FCT-MEC in the framework of the  PEst-OE/ EEI/UI0527/2014 strategic project.}

\author{Fábio Soldado}

\author{Fernando Alexandre}

\author{Herv\'e Paulino}
 \ead{herve.paulino@fct.unl.pt}
 \ead[url]{http://asc.di.fct.unl.pt/~herve}

\address{NOVA Laboratory for Computer Science and Informatics \&  Departamento de Informática, Faculdade de Ciências e Tecnologia, Universidade Nova de Lisboa, 2829-516 Caparica, Portugal}

\begin{abstract}
Current computational systems are heterogeneous by nature, featuring a combination of CPUs and  GPUs.
As the latter are becoming an established platform for high-performance computing, 
the focus is shifting towards the seamless programming of these hybrid systems as a whole.
The distinct nature of the architectural and execution models in place raises several challenges, as the best hardware configuration
is  behaviour and workload dependent.
In this paper, we address the  execution of compound, multi-kernel, OpenCL computations in multi-CPU/multi-GPU environments.
We address how these computations may be efficiently scheduled onto the target hardware, 
and how the system may adapt itself to changes in the workload to process and to fluctuations in the CPU's load.
An experimental evaluation attests the performance gains obtained
 by the conjoined use of the CPU and GPU devices, when compared to GPU-only executions, and also
by the use of data-locality  optimizations in CPU environments. 
\end{abstract}


\end{frontmatter}

\section{Introduction}

Most of the current computational systems are intrinsically heterogeneous, featuring a combination of  multi-core CPUs and  Graphics Processing Units (GPUs). 
However, the discrepancies of the programming and execution models in place
make the programming of these hybrid systems a complex chore, 
as several issues must be addressed:
which computations to run on each kind of processing unit; 
how to decompose a problem to fit the execution model of the target processing unit; 
how to map this decomposition in the system's complex memory hierarchy; to name a few. 
Consequently, only experts with  deep knowledge of parallel programming, and even computer architecture,
are able to fully harness the available computational power.

The OpenCL specification has been designed  with the purpose of enabling code portability across a wide range of architectures.
However, performance portability  is not guaranteed.
In fact, it  greatly depends on device-specific optimizations, which are
cumbersome to implement, due to the low level nature of the  programming model.
Moreover, when targeting multiple devices, it is still up to the programmer to  decompose  the problem for multi-device execution.
These limitations have been driving 
a considerable amount of  research  in the field of heterogeneous  computing.
A growing trend is to build upon the notions of algorithmic skeletons and templates.
This is mostly visible at library level \cite{skepuadaptative,skelcl_multi,bolt,thrust,muesli},  but languages such as StreamIt \cite{streamit-gpu}  and Lime \cite{lime}  also apply this type of constructions to GPU computing.

We share this vision. Algorithmic skeletons render a programming model that abstracts the complexity inherent to parallel programs by factorizing known solutions in the  field into high level parameterizable and composable structures.
 We claim that these characteristics can be used to: (a) hide the heterogeneity of the underlying hardware and, (b) provide tools to cope with such heterogeneity, enabling device-specific problem decompositions and optimizations. 
To that extent, we have been developing the  Marrow algorithmic skeleton framework  \cite{marrow,marrow_sac,marrow_hetero} 
for the orchestration of OpenCL computations. Marrow offers both data and task-parallel skeletons 
 and is the first framework  on the GPU computing field to support skeleton composition, through nesting. 
Marrow computations are thus  compositions of multiple OpenCL kernels.

In this paper, we focus on how to execute 
Marrow  computations
on multi-CPU/multi-GPU environments and, with that, offer
a skeletal  programming model 
for the transparent programming of hybrid  CPU/GPU systems.
To that end, we  equipped the Marrow framework with optimized back-ends for  the execution of OpenCL computations on both CPUs and GPUs, and automatically tune the framework, so that it can efficiently execute a given computation among 
the selected hardware.
Tuning the framework requires the  derivation of
   a set of  framework configuration parameters 
(for both CPU and GPU-directed executions), and 
of
 a  workload distribution among the multiple processing units.
We allow the  process to be conducted in advance, before the beginning 
of the computations, or  online, as   computations  are executed.

The work distinguishes itself from the current state of the art by supporting the execution of arbitrary multi-kernel compound computations, having in mind data locality requirements.
The current state of the art  either exposes the heterogeneity to the programmer \cite{chapel-gpu,muesli} or
 selectively directs the computations exclusively to one of  the available CPU or GPU back-ends  \cite{skepuadaptative,skelcl_multi,bolt,thrust,adaptativegpu,dandelion}. 
In turn, the proposals that tackle the  transparent 
conjoint use of both CPUs and GPUs either restrict their scope to the execution of single kernels
 \cite{mapreducecpugpu,LeeSPM13,qilin} or 
 require previous knowledge on the computation to run \cite{tecnicoppam}.

The contributions of this paper are thus:
\begin{inparaenum}[a)]
\item a locality-aware  decomposition of  Marrow compound computations among  multiple  devices, 
\item strategies to distribute the load of a Marrow   computation among  multiple CPU and GPU devices, adapting
 this  distribution to different  workloads and to the CPUs' load fluctuations,
\item the seamless integration of these concepts in the Marrow programming model,
\item an experimental evaluation that demonstrates the usefulness of the integration of multi-CPU support in both CPU-only and hybrid CPU/GPU environments.
\end{inparaenum} 

The remainder of this paper is organized as follows: 
the next section gives a general overview of the Marrow skeleton framework;
Section \ref{sec:cpu_gpu} describes our strategies for the cooperative multi-CPU/multi-GPU execution
of Marrow  compound computations;
Section \ref{sec:eval} experimentally evaluates our proposal from a performance perspective;
Section \ref{sec:related} compares our approach to the current state of the art; and, finally, Section \ref{sec:conclusions} presents our final conclusions and prospective future work.
\section{An Overview of  the Marrow Algorithmic Framework}
\label{sec:skel_lib}

\begin{figure}
\begin{floatrow}
\ffigbox[8cm][4.8cm]{
 \centering
  \includegraphics[height=3.8cm]{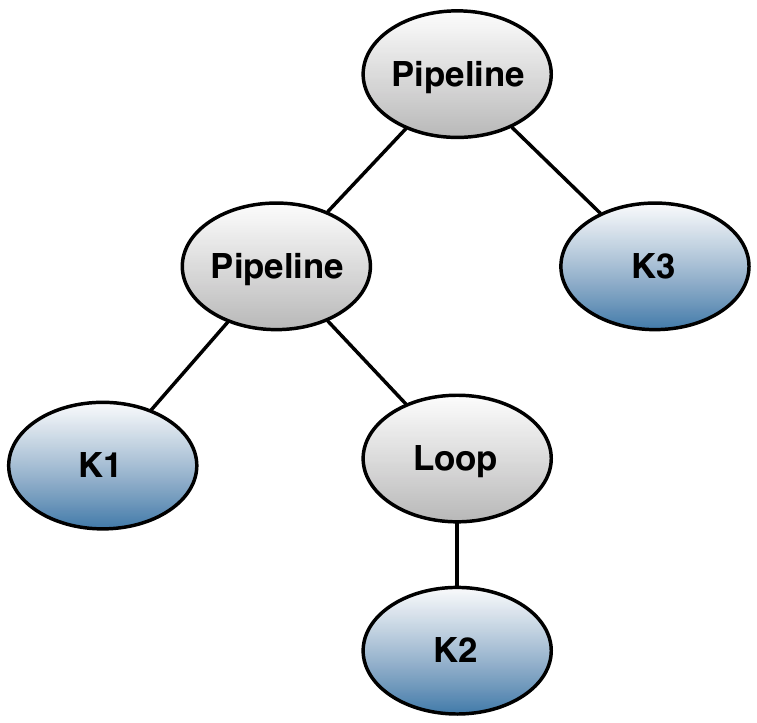}
}
{
  \caption{A Marrow SCT: \textsl{pipeline}($K_1$, \textsl{loop}($K_2$), $K_3$)}
  \label{fig:ct}
}
\ffigbox[4.5cm][4.8cm]{%
    \centering
    \includegraphics[height=4cm]{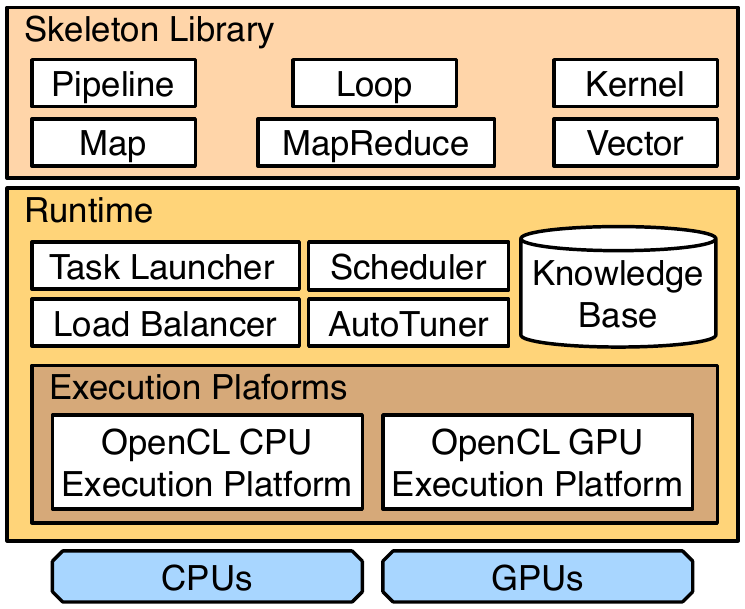}
}{
  \caption{Marrow's architecture}
  \label{fig:arq-bitmap}
}
\end{floatrow}
\end{figure}

Marrow is a C++ algorithmic skeleton framework for the orchestration of OpenCL  computations.
It provides a set of data and task-parallel skeletons that can be combined, through nesting, to build compound behaviors.
A Marrow computation may is thus   a tree of skeleton constructions (Fig. \ref{fig:ct}), each applying   
  a specific behavior 
to its sub-tree, down to the leaf nodes -  the actual OpenCL kernel computations.
The  framework takes upon itself  the entire host-side orchestration required to correctly execute these skeleton computational trees (SCTs) in multi-CPU/multi-GPU environments, including the proper ordering of the
data-transfer and execution requests, and the communication  between the tree nodes.

\label{sec:marrow:exec_model}
Marrow's execution model is directed at  batch computations.
Execution requests are handled according to a first-come-first-served policy, being that each SCT execution makes use of  all the hardware made available to the framework.
These requests may target one or more SCTs, given that, once built, a SCT may receive multiple execution requests, over possibly distinct data-sets (in both content and size).
Restricting, for now, our scope to single device executions, 
	the kernels encased  in a SCT (which may be a sub-tree of the entire submitted computation)
are executed sequentially,  according to a depth-first evaluation of the tree.
For example, the kernel execution  order  for the SCT depicted in Fig. \ref{fig:ct} is:  ${K_1}$, followed by multiple executions of  $K_2$, and lastly  $K_3$.

In what concerns its architecture, the framework
follows a two layered structure:  the upper \textsl{Library} layer includes all the components
necessary for the programming of Marrow SCTs, 
whilst the lower \textsl{Runtime} layer provides for the execution of such SCTs in multi-CPU/multi-GPU architectures.
Fig. \ref{fig:arq-bitmap} presents an overall view of both layers, whose components we briefly describe below.

\subsection{The Library}
The Marrow library offers the following set of skeletons:
\begin{inparadesc}
\item [\texttt{Pipeline}] - a pipeline of control and data-dependent SCTs;
%
%\item[Seq] - a pipeline of an arbitrary number of control-dependent CTs, no data is implicitly exchanged between the CTs;
%
 \item[\texttt{Loop}] -  \textit{while} and \textit{for} loops over a SCT;
\item[\texttt{Map}]  - application of a SCT upon independent partitions of the input data-set, and;
 \item[\texttt{MapReduce}] - extension of  \texttt{Map} with a subsequent reduction stage.
\end{inparadesc}

A SCT may embed  multiple arbitrary OpenCL kernels.
These take the form of  \texttt{Kernel} objects, which  enclose the  kernel's logic and domain in a single computational unit.
Given the arbitrary nature of the computations, for the sake of correctness and efficiency,  setting up 
\texttt{Kernel} object requires the specification of the interface  of the wrapped computational kernel, namely in what concerns its input and output parameters.
For that purpose,  the library supplies a set of  data-types to, on one hand, classify these parameters
 as vector or scalar values, 
and, on the other, to express if they are immutable values and whether they must be allocated in local memory.
Moreover, the programmer may supply a kernel-specific  work-group size for computations that are bound to particular sizes, 
and  specify upon how many elements of the multi-dimensional range each computing thread (aka OpenCL work-unit) operates on (defaulted to 1).
For instance, a single thread  may work upon multiple pixels of an image. 
This information is used to compute the number of threads (OpenCL work space) required to run the kernel.

\begin{table}
\resizebox{\textwidth}{!}{
\begin{tabular}{l}
\hline
\textbf{SCT construction:}\\
\quad \texttt{OpenCLKernel} \textbf{kernel} (\texttt{std::string kernel\_file, std::string kernel\_function, std::vector<IDataType> input\_args}, \\ 
\qquad \texttt{std::vector<IWorkData> output\_args, [std::vector<int> local\_work\_size], [int work\_per\_thread}]) \\
\quad \texttt{Pipeline} \textbf{pipeline}(\texttt{SCT... stages}) \\
\quad \texttt{Loop} \textbf{loop}(\texttt{SCT body, LoopState state})\\
\quad \texttt{Map} \textbf{map}(\texttt{SCT  tree})\\
\quad \texttt{MapReduce} \textbf{map\_reduce}(\texttt{SCT  map\_stage}, \texttt{SCT  reduction\_stage})\\
\quad \texttt{template $<$class R, class... Args$>$ MapReduce} \textbf{map\_reduce}(\texttt{SCT  map\_stage}, \texttt{std::function$<$R, Args$>$  reduction\_stage})\\
\\
\textbf{Execution request: }\\
\quad \texttt{template $<$class R, class... Args$>$ Future$<$R$>$} \textbf{run}(\texttt{Args... arguments}) \\
\hline
\end{tabular} 
}

\begin{scriptsize}
\begin{tabular}{p{\textwidth}}
 \texttt{IDataType} denotes a kernel argument type. It is
 subclassed by \texttt{VectorType<T>} and
\texttt{ScalarType<T>} (each with mutable and immutable variants) to express
vectors and scalars, respectively. 
\texttt{SCT} denotes the interface that must
be implemented by every Marrow tree element.
\end{tabular}
\end{scriptsize}\\
\hrule
\caption{Marrow constructs - optional arguments are delimited by 
square brackets.}
\label{tab:api}
\end{table}

The setting up of the computation tree is a bottom-up process. 
Each skeleton construct takes the list 
of skeletons and/or kernels
necessary to instantiate its abstract behavior, plus a set of skeleton specific parameters.
\texttt{Loop}, for instance, requires the specification of a state comprising 
the loop's stoppage condition, 
the data items that must be updated after 
each iteration,  and  whether this update operation requires global (all device) synchronization.  
Table \ref{tab:api} presents Marrow's key programming constructs.

Execution requests may be issued upon any node of an SCT, requiring only 
 the  list of  object(s) in the host's memory that must be read and/or written by the SCT.
The operation is asynchronous, returning a future object (Table \ref{tab:api}).
The arguments  may be scalar values or \texttt{Vector} data containers.
\texttt{Vector}  exposes an interface similar  to the one of \texttt{std::vector}
and abstracts all data management operations, such as localization and transfers, being relatively close  to  what may be  found in \cite{skepuadaptative,skelcl_multi,muesli,bolt}.

\subsection{The Runtime}
\label{sec:lib:exec_model}
The \textsl{Runtime}  is itself subdivided into two inner layers.
The upper one congregates all the functionalities  that are independent of the technology
used to achieve the desired behaviors on the target processing units.
Its main modules are:
\begin{inparaitem}
\item[\textsl{Scheduler} -] distributes the  execution of an SCT  among the selected hardware, generating a group of tasks 
that  are placed in a set of work queues (one per parallel execution, being that there may be multiple parallel executions per device (Section \ref{sec:profile_const}));
\item[\textsl{Task Launcher} -] consumes the tasks from the work queues and launches their execution in the target execution platforms;
\item[\textsl{Load Balancer} -] redistributes the load of the current SCT among the multiple CPUs and GPUs.
\item[\textsl{Auto Tuner} -]  finds a suitable framework configuration for the execution of a given SCT;
\item[\textsl{Knowledge Base} -] a database that stores information about the configuration settings  of past executions, plus
a inference engine able to deduce configurations for newly arriving SCTs.
\end{inparaitem}

The multi-layer design relegates to the lower layer all technology-bound implementation details, such as host-device communication and device-specific optimizations.
The approach promotes the use (or even combination) of multiple back-ends, baptized as \textsl{execution platforms}.
In this paper our focus is directed exclusively to OpenCL, for which we have implemented two platforms to cope with the specificities of
both CPUs and GPUs. 
The \textsl{GPUExecutionPlatform} applies  multi-buffering techniques to support the overlap of computation with communication in GPU-directed  executions, while the \textsl{CPUExecutionPlatform} uses
the OpenCL device fission functionality \cite{opencl_fission} to leverage data locality in CPU-directed executions.
The latter enables a memory-hierarchy aware partitioning of  OpenCL CPU devices comprising one or more multi-core processors.

\section{Cooperative Multi-CPU/Multi-GPU Execution}	
\label{sec:cpu_gpu}

The  contribution of this paper is on the adaptable execution of Marrow SCTs in hybrid multi-CPU/multi-GPU
environments.
Our goal is to offer a single-image view of the system that abstracts the heterogeneous nature of the underlying hardware platform.
To accomplish such  enterprise we must address four key challenges that will drive the structural organization of this section:
\begin{inparaenum}[a)]
\item how to efficiently decompose a SCT among  multiple CPU and GPU devices;
\item how to efficiently distribute a workload among the available hardware resources;
\item how to dynamically adapt this  distribution to changes in the  workload and to fluctuations of the CPUs' load;
\item how to  integrate these concepts in the Marrow programming model in a non-intrusive way.
\end{inparaenum}

Marrow's  modular architecture  allows us to concentrate  our efforts 
in the \textsl{Runtime} layer. 
The impact of multi-device execution in the \textsl{Library}  is minimum and will be discussed in Section \ref{sec:cpu_gpu:programming}.

\subsection{Skeleton Computational Tree Decomposition}
\label{sec:decomp}

We address this challenge  by leveraging on the work-group based organization of OpenCL executions, namely on the fact that work-groups (groups of threads) execute asynchronously and  
 independently over data.
%the fact that OpenCL kernel synchronization is confined to work-groups.
Consequently,  we opt to decompose the computation's data-set into partitions that can be adjusted to the best possible work-group size for each device. % (see Section \ref{}).
In that sense, we extend the scope of OpenCL's SPMD (Single Program Multiple Data) based execution model to multiple devices, where each OpenCL work-group computes the SCT over a partition of the input data-set according to 
the single device execution model described in Section \label{sec:marrow:exec_model}.

An alternative decomposition approach would be to dismantle the SCT across the multiple devices.
The procedure promotes  the concurrent execution of different kernels (leaf nodes)
but leads to the data movement between devices, making the system more sensible to the latency of data transfers along the PCIe bus.
The workload distribution process would have to be aware of  
the amount of data to be transferred among nodes of the SCT,  
as well as of the number of PCIe-attached devices assigned to the tree's execution.
There is a inherent trade-off between moving data among	 running computations or moving computations towards data.
The latter adapts better to our  focus on batch computations.
However, future incursions in streaming support may  shift this balance towards the former,
as kernels  will concurrently process streams of data flowing up and down the SCT.

\begin{figure}
\centering
\includegraphics[height=3.5cm]{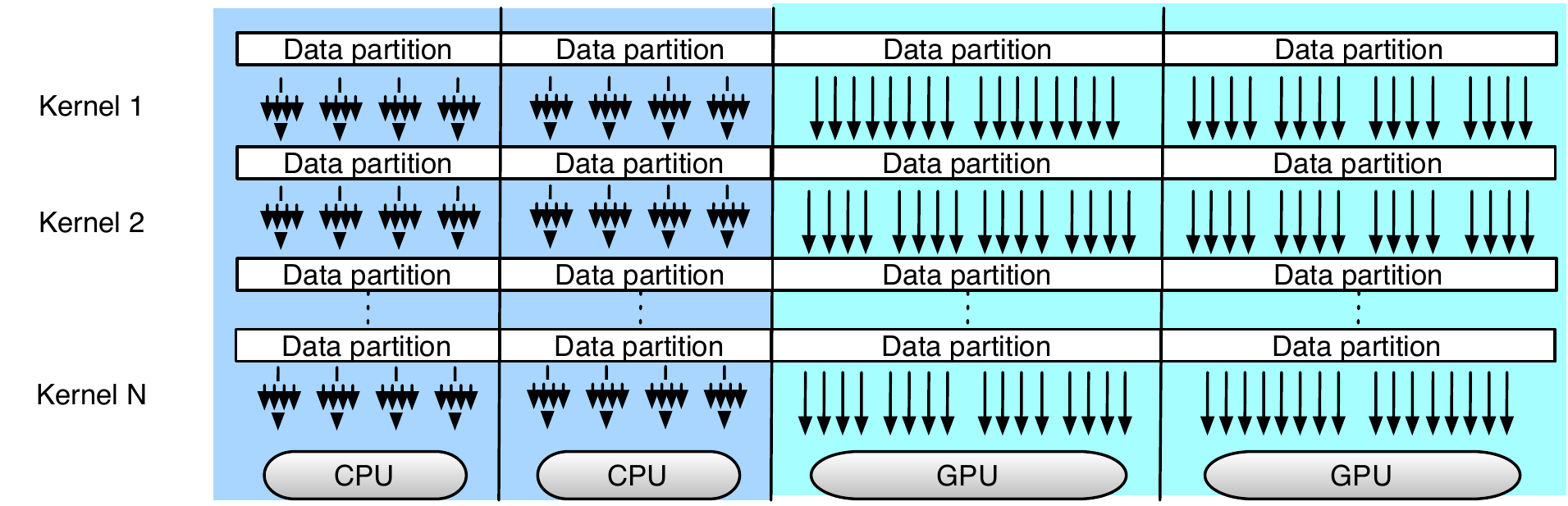}
\caption{Execution of a SCT across multiple CPU/GPU devices}
\label{fig:multi-device}

\end{figure}

Coming back to the implemented SPMD-based execution model, 
the partitioning of  input data-sets must be subjected to  the user-defined restrictions conveyed in the target SCT's  specification, and  be driven by the characteristics of the underlying hardware, such as the size of AMD \textit{wavefronts} or NVIDIA \textit{warps} in GPUs.
Furthermore, in order to really leverage the aforementioned locality properties,  
the communication of 
data  between two consecutive kernel executions must be achieved via the simple  persisting  of this data in the devices' memory, i.e. without requiring data movement between devices.
As a result, two kernel executions that communicate one or more data-sets must 
expect an identical  partitioning of such sets, with respect to their number and size(s),
regardless the individual work-group size restrictions of either kernel.
This approach induces a partitioning process with  a global vision of the computation, that  we have named \emph{locality-aware domain decomposition}.
Fig. \ref{fig:multi-device} illustrates a simplified view of our execution model.
The  arrows denote the OpenCL threads, which 
in the CPU devices are distinguished between the ones support by hardware threads (the long arrows)
and the ones supported by the vectorization units (the smaller ones). 
The number of threads and their organization in work-groups may differ between kernel-executions.

The locality-aware domain decomposition process takes into consideration
the arguments passed to the SCT's execution request,  
the interface specified for each kernel in the SCT,
and the occupancy of each of these kernels in the target GPU device(s).
Kernel occupancy is computed according to the usual constraining factors \cite{amd_kernel_occupancy}: 
the number of work-groups per compute unit, 
the amount of local memory per work-group, and
the number of general purpose registers (private memory) required by each thread within a group.
In turn, multi-device support
 adds new ingredients to  kernel  specification -  information must be conveyed about which input vectors may, or   not,  be partitioned among the devices, and, in the former case,
which is the \emph{elementary partitioning unit}, in number of elements. 

Given this kernel interface specification framework, assume  the existence of two arbitrary kernels $K_1$ and $K_2$ that communicate via a set of vectors $V_1, V_2, \dots, V_n$.
Each of these vectors must be partitioned in the same number of partitions, $p$;  value ruled by the number of parallel executions (of the SCT).
However, this partitioning process must respect a set of constrains.
We thereby define the following notation: $V_i^j$ to denote the partition of vector $V_i$ associated to execution $j$;
 \textsl{wgs}$_j$($K$) to denote a valid   work-group size specified for kernel $K$ on the device running execution $j$; \textsl{epu}($V$)  to denote $V$'s  elementary partitioning unit
 and  \textsl{nu}($V, K$) to  denote  the number of elements of $V$ computed by each thread of $K$.
 Accordingly,  the constraints to be applied upon the partitioning of a SCT may be formulated as follows:

\begin{small}
\begin{eqnarray*}
\textstyle \forall\ i, j: V_i = \bigcup_{j=0}^p V_i^j 
&&    \textsl{epu}(V_i)\ \text{mod}\ \textsl{nu}(V_i, K_1) = 0 \quad \text{and} \quad  \#V_i^j\ \text{mod}\ (\textsl{epu}(V_i)  / \textsl{nu}(V_i, K_1)) = 0 \\
&& \textsl{epu}(V_i)\ \text{mod}\ \textsl{nu}(V_i, K_2) = 0 \quad \text{and} \quad  \#V_i^j\ \text{mod}\ (\textsl{epu}(V_i) /  \textsl{nu}(V_i, K_2)) = 0  \\
%\text{(as much as possible)} 
&& \#V_i^j\ \text{mod}\ \textsl{wgs}_j(K_1) = 0 \quad \text{and} \quad  \#V_i^j\ \text{mod}\ \textsl{wgs}_j(K_2) = 0 \\
\end{eqnarray*}
\end{small}

Not all skeletons comply entirely to the cross-device SPMD model.
The \textit{Loop} and \textit{MapReduce} skeletons delegate parts of their execution exclusively to the host.
A Marrow \textit{Loop} comprises  three stages: 1 - evaluation of the condition, on the host; 2 - execution of the body (the SCT), on the device(s); and 3 - the update of the loop's state according to 
the memory positions written by the SCT (also performed on the host).
This last stage is carried out in parallel whenever the state's updating code may be applied independently to the result of  each partial execution.
In the \textit{MapReduce} case, given the difficulty of implementing efficient reductions on GPUs, the skeleton also accepts C++ functions that are   executed on the host side. It is thus up to the programmer to decide \emph{where} the reduction takes place.

\subsection{Workload Distribution}   
\label{sec:hw_selection}

In order to efficiently execute compositions of arbitrary OpenCL kernels in hybrid multi-CPU/multi-GPU environments,
we must engender a solution that is able to deliver good performances without requiring previous knowledge on the SCT to execute.
Concerning GPUs, the workload is statically distributed among the devices, according to their relative performance.
We establish this order relation for both integer and floating-point arithmetic by running the SHOC benchmark suite \cite{shoc} at the framework's installation-time.
This static approach, although simple, delivers good performance results for GPU-accelerated executions \cite{marrow_sac}, due to the specialized nature of the underlying execution model:
one kernel execution at a time, with no preemption and no input/output operations.

These premises are not valid  for CPU executions.
The execution time of a CPU computation is highly conditioned by the load of the processor (time-shared by multiple threads), and by  hardware optimizations that cannot be fully controlled by the programmer, such as   cache  memory management.
Therefore, to adequately schedule arbitrary computations between CPU and GPU devices is still a challenge.

\begin{figure*}
\includegraphics[width=\linewidth]{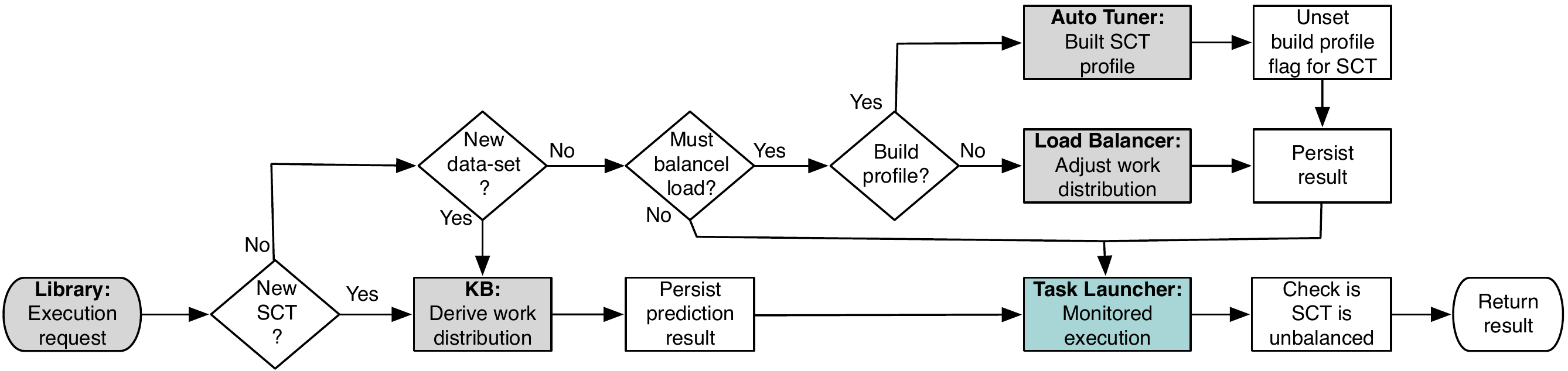}
\caption{The top-level work-distribution decision process}
\label{fig:process}

\end{figure*}

In the context of OpenCL computing, this scheduling procedure must determine an efficient   \textit{distribution of work}  among the devices,
and, for each of such devices, ascertain the best \textit{device-specific configuration}  for the pair (computation, data-set partition(s)).
The calculation of these multiple framework configuration parameters is time and resource consuming.
To that extent,  we researched how to leverage 
the knowledge accumulated in time, from the execution of many different SCTs over distinct workloads,
in order to derive a suitable  configuration for the execution of an arbitrary SCT upon an arbitrary  data-set.
We address the challenge via profile-based self-adaptation.
We still rely on static scheduling: the workload is distributed in advance between the selected devices. 
However, this distribution process resorts to  profiles built from past runs.
Furthermore,  we continuously refine this information, so that 
subsequent executions may better adapt to fluctuations of the CPUs' load 
and/or to changes in the SCTs' workload.
For that purpose, SCT executions  are monitored by a controlling process
that  identifies, and corrects,  load unbalances.
To be noted that changes on the workload do not include changes in the actual values being computed, but only on load's characteristics, such as the number of elements.

The main workflow of the workload distribution process (depicted in Fig. \ref{fig:process})
is executed by  the \textsl{Scheduler} module, interacting with  the remainder where highlighted in the figure.
Upon the reception of a new execution request from the \textsl{Library}, if either the target  SCT or the submitted workload 
are different from the ones  executed on the previous run, 
a new framework configuration has to be 
\textit{derived}  (box "\textit{Derive work distribution}"). 
Central to this decision process is the  availability of the \textsl{Knowledge Base} (KB) 
that stores information about the best known configuration  for a given pair (SCT, workload).
It is always assumed that the KB  holds enough information to supply a  configuration
able to deliver good performances. If such assumption reveals itself as too optimistic, 
adjustments to workload distribution will be made, and the derived profile refined.

%-----  (see Section \ref{sec:dynamic}).
%
When in the presence of a recurrent SCT execution upon the same workload,  
the process proceeds to assess if the previous runs of such computation  were unbalanced.
If needed be, it triggers the dynamic load balancing mechanism to adjust the current workload distribution
(box "\textit{Adjust workload distribution}"), or 
simply builds a SCT-specific profile from scratch, if such profile does not exist (box "\textit{Build SCT profile}"). 
Note that both the derivation and profiling  branches conclude with the persisting of the attained result in the KB,
qualified with the process employed. 
As a result, the derivation process also contributes to populate the KB, serving as a cache for following executions.

\subsubsection{Profile Specification:}
A profile of a Marrow SCT  contains all the information necessary to reproduce a configuration of the framework, namely:
\begin{inparaenum}[a)]
\item an SCT unique identifier;
\item a workload characterization, namely its number of dimensions,  number of elements per dimension, and if it  includes single or double precision floating-point data;
\item the percentage of the workload to be assigned to each device, being it a CPU, a GPU or any other to be supported in the future;
\item  the  configuration of the  execution platform associated to  each of these devices;
\item the minimum execution time measured for the stored configuration - useful for later refinements, 
and, lastly;
\item the profile generation process, one of: derived from the KB or built from empirical data.
\end{inparaenum}

\subsubsection{Profile Construction (box \textit{"Build SCT profile"}):}
\label{sec:profile_const}
The construction of a profile from scratch is performed only once for each pair (SCT, workload), 
and is triggered only if two conditions hold: i) the current configuration is not generating
 well-enough balanced executions across the multiple CPU and GPU devices, and ii) the framework was explicitly configured to branch to this profile building alternative.
The profiling  mechanism is specially tailored for applications that operate over similar  workloads for long periods of time.
In such scenarios, it compensates to have the best possible configuration, even if it takes some time to obtain it.

The profile construction process searches for the best workload distribution and execution platform configuration for 
the computation at hand.
The \textsl{CPUExecutionPlatform} supplies an iterator over the affinity fission configurations the device supports\footnote{For the sake of readability, in this document we restrict ourselves to partitioning by affinity domain, although we also have experimented with partitioning by equality.}, a subset of:
${\textstyle (\bigcup_{i = 1}^4 \{L_i\_CACHE\}) \cup \{NUMA\} \cup \{NO\_FISSION\}}$.
In turn, the \textsl{GPUExecutionPlatform }
offers an iterator over  two ordered sets:  i)
a list of work-group size values that ensure (if possible\footnote{Otherwise the work-group size that yields the best occupancy is selected}) a device occupancy above a configurable threshold (defaulted to 80\%),
and ii) $[1, \infty] \subset \mathbb{N}$, representing the number of overlapped executions to be performed by the GPU(s).

The profiling process treats the multiple CPUs and GPUs as indivisible units. 
We are assuming conventional homogeneous multi-core CPUs, and, %multi-CPU systems are inherently homogeneous.
in what concerns heterogeneous multi-GPU systems, we perform the static work distribution described  in Section \ref{sec:decomp}.
Note, however,  that both the use of CPU fission and GPU overlapped executions contribute to the increase of the number of data-set partitions.
CPU fission actually divides the original multi-CPU OpenCL device into smaller devices, and
the use of \textit{overlap} overlaps the computation of a SCT upon a partition with the transfer of (some of) the remainder partitions assigned to that same GPU.
Thus, taking another look at  Fig. \ref{fig:multi-device}, each device may be responsible for  multiple concurrent executions.
The grand total of these executions defines in what we entitle as the SCT's  \emph{level of (coarse) parallelism}, which is combined with the fine grained parallelism of inner work-group execution.

The quest for the best possible  overall performance for the submitted (SCT, workload) pair 
requires finding  the globally best performing tuple (CPU fission level, GPU overlap, Per kernel work-group size,  CPU/GPU workload distribution).
Algorithm \ref{alg:wld} presents the main steps of the proposed procedure.
Initially, the entire configuration search space is retrieved  from both execution platforms (steps 1 to 3).
The  fission level and overlap values are driven only by  hardware characteristics, but the validity of the  work-group size candidate values 
depends on both the hardware and the computation. 
Ergo, for generality's sake, information about the computation is always made available to the \texttt{getConfigurations} function.
%
%The setup stage concludes in step 4 with the setting-up of the load distribution generator, 
%which will be responsible for proposing load distributions between CPUs and GPUs, given a particular framework configuration.
%%

The next step is to iterate over the possible execution platform configurations.
Given that the search space is quite large, we do not test every possible solution.
The multiple dimensions of the search space are ordered according to the likeliness of the candidate values to deliver good performances, and the search is limited by a stoppage criteria 
defined in the \texttt{discard} function
installed in every configuration-storing object (steps 21, 23 and 25). 
Accordingly,  CPU fission levels are ordered from $L1$ to $NO\_FISSION$, 
the GPU overlap factors are ordered according to the natural order, and 
the GPU work-group sizes are ordered in non-increasing order of  GPU occupancy.
In all cases, whenever a candidate value 
 fails to improve  performance relatively to the former, all subsequent ones are discarded.

\begin{algorithm}[t]
\hrule
\begin{tiny}
\caption{Profile Building Algorithm}
\label{alg:wld}
\LinesNumbered
\KwIn{\texttt{SCT} - the skeleton computation tree} 
\KwIn{\texttt{arguments} - the  array of the SCT's input arguments} 
\KwIn{\texttt{SCT\_id} - the SCT's unique identifier} 
\KwIn{\texttt{workload\_id} - the workload's characterization} 
\KwIn{\texttt{occupancy\_threshold} - minimum accepted occupancy} 
\KwIn{\texttt{precision} - precision value for the workload distribution} 
\KwIn{\texttt{number\_executions} - quality value for workload distribution} 
% $best\_time \leftarrow \infty;$ \\ 
$cpu\_configurations \leftarrow $ CPUExecutionPlaform.getConfigurations(\texttt{SCT}, \texttt{arguments}) // \{ fission levels \} \\
$gpu\_configurations \leftarrow $ GPUExecutionPlaform.getConfigurations(\texttt{SCT}, \texttt{arguments}) // \{ \{  overlap\_factors \}, \{ work-group sizes \} \}\\
$workgroup\_sizes \leftarrow$ filter($gpu\_configurations$.get(1), \texttt{occupancy\_threshold}); \\
\For{$fission\_level \in  cpu\_configurations$.get(0)}{
  $cpu\_par\_level \leftarrow$ CPUExecutionPlaform.configure($fission\_level$);\\
 % \lIf{cpu\_par\_level = -1}{\textbf{continue}} 
    \For{$overlap \in gpu\_configurations$.get(0)}{
      $gpu\_par\_level \leftarrow$ GPUExecutionPlaform.configure($overlap$);\\
      \For{$wg\_sizes \in workgroup\_sizes$}{
       $WLDG \leftarrow$ workload\_distribution\_generator($cpu\_par\_level, gpu\_par\_level$);\\
      \While{true}{   
            $dist \leftarrow WLDG$.next(); \\
            $data\_partitions \leftarrow$  partition\_input(\texttt{SCT}, \texttt{arguments}, $dist, wg\_sizes$); \\
           $time \leftarrow$ exec\_for\_profile(\texttt{SCT}, $data\_partitions$, $wg\_sizes, WLDG$, \texttt{number\_executions}); \\
           $best\_time \leftarrow $ get\_stored\_profile().best\_time \\
           \If{$time < best\_time$}{
              store\_profile(\texttt{SCT\_id}, \texttt{workload\_id}, $fission\_level, overlap, wg\_sizes, dist, time$); \\
              \lIf{$(best\_time - time) < \texttt{precision}$}{\textbf{break}}
           }
           \lElse {\textbf{break}}
       }
         \lIf {$gpu\_configurations$.discard($wg\_sizes$, get\_stored\_profile())} {\textbf{break}}
      }
 	   \lIf {$gpu\_configurations$.discard($overlap$, get\_stored\_profile())} {\textbf{break}}
  }
  \lIf {$cpu\_configurations$.discard($fission\_level$, get\_stored\_profile())} {\textbf{break}}
}
\end{tiny}
\hrule
\end{algorithm}

For each possible (CPU fission, GPU overlap, work-group sizes) configuration, the algorithm
first  reconfigures the  target execution platforms  (steps 5 and 7), so that 
the framework yields the established level of  parallelism.
Secondly, it computes the best workload distribution for the given configuration (steps 9 to 20).
To carry out this enterprise, the innermost loop performs five steps:
\begin{inparaenum}
\item[step 11 -] obtain the next candidate workload distribution ($dist$) from the workload generator (detailed ahead);
\item[step 12 -]  partition the input data-set according to $dist$, but also taking into consideration the 
work-group sizes candidates, and the kernels' elementary partitioning units (embedded in the \texttt{SCT});
\item[step 13 -]  execute the \texttt{SCT} given the current configuration and workload distribution  (\texttt{number\_executions}
serves as a quality factor to avoid possible performance fluctuations);
\item[steps 15 and 16 -]  check if the latest execution is the one that  yielded the best performance thus far, and, if so, store 
all configuration parameters in the profiling information, and, lastly;
\item[step 17 -] conclude the current search direction, if the difference between the execution times of two consecutive overall configurations
is smaller than the precision value.
\end{inparaenum}

The \textit{workload distribution generator} acts like an iterator that, at each invocation, 
outputs a CPU/GPU distribution that tries to even, as much as possible, the  time 
that each device type (CPU or GPU) takes to carry out the SCT.
To that end it  performs a  binary search that, at each iteration transfers load from the worst to the best performing device-type.
The strategy  requires the tracking of the amount of work that can be transferred between the device types (the \textit{transferable partition}), and of the amount that is already bound to each type.
Initially, all the  work is considered to be transferable and, hence, none  of it is bound to any device type.
With each iteration, the transferable partition is evenly split  between the two device types, 
and permanently bound to the one that performed better.
The remainder half will become  the next transferable partition, which 
may be regarded as the portion of the work that is still under training. 
Its size after $n$ iterations is given by function  $\textsl{transferableSize}(n, size) = \frac{size}{2^n} 
\text{ with asymptotic behaviour } \lim_{n \to \infty} \textsl{transferableSize}(n, size)  = 0$.
Thus, the impact of each training iteration in the resultant work distribution will likely decrease with each iteration.
Given the need for the partitions to comply to the restrictions imposed by the SCT specification and  the characteristics of the underlying hardware, performance fluctuations between devices may arise.
When so happens, distribution fairness is not always in hand with the best performance possible.
As such, the solution found may be inherently unbalanced.

 \subsubsection{Configuration Derivation (box  "Derive work distribution"):}

The configuration derivation process is triggered whenever  there are alterations on the  (SCT, workload) pair to process.
The result must provide the workload distribution for all the CPU and GPU devices, as well
as configurations for the execution platforms.
The  enterprise is trivial if the necessary information is already available in the KB, otherwise it will have to be derived from the previously collected knowledge.
For that  purpose we apply multidimensional interpolation algorithms for scattered data -  the number of dimensions is
directly associated to the dimensionality of the computation's workspace. 
For dimensions 1 to 3, we apply the Fast Radial Basis Function Network algorithm provided by the Alglib numerical analysis library\footnote{\url{http://www.alglib.net}},
whilst for dimensions $> 3$ we employ a \emph{nearest-neighbour} method sustained by the Euclidean distance.
%
%The framework allows for the simple integration of new interpolation methods, and thus others may be incorporated.

The scope of the interpolation algorithm is initially narrowed to  the  previously collected configurations  for the target SCT.
 If no such information exists, the algorithm is then applied to configurations available for the submitted workload ($w$), disregarding the SCT.
 Lastly, if once again  no information could be retrieved, it is  applied to all workloads of the same dimensionality of $w$.

 \subsection{Dynamic Load Balancing}

Every SCT execution is monitored with the objective of generating  a  set of  useful statistics, among which are the time required to complete each concurrent execution of the SCT over a partition of the supplied data-set,
the deviation (${dev}$) between these times,
%$$\textstyle performance = \sfrac{shortestDuration}{longestDuration}$$
and the load balancing threshold (${\textsl{lbt}}$)  for  the current execution ($n$). 
The latter, defined as follows, establishes the threshold over which a SCT's execution is considered to be unbalanced:

\begin{eqnarray*}
 \textsl{lbt}(n) & = & \textsl{isUnbalanced}({dev}) \times {weight} + {\textsl{lbt}}(n-1) \times (1-{weight}) \\
\text{for} \quad
\textsl{isUnbalanced}(x)  &= & \left \{ 
\begin{array}{l l}
0 & \text{if}\ \frac{x}{{cFactor}} \leq {maxDev} \\
1 & \text{otherwise} 
\end{array} \right. 
\end{eqnarray*}

%}

\noindent
where ${weight}$ denotes the weight assigned to the last execution relatively to the historical data, 
${maxDev}$ is a user-definable upper bound for the execution's deviation, in order to be considered balanced, 
and  ${cFactor}$ is a correction factor to deal with computations that perform better with slightly 
unbalanced work distributions (see Section \ref{sec:profile_const}).
A SCT is considered to be unbalanced  when  $\textsl{lbt}(n) \approx  1$.
The use of a weighted historical data factor makes the result of $\textsl{lbt}$ less sensitive to sporadic unbalanced executions.
%Table \ref{tab:lbt} illustrates its evolution 
For the framework's default  $weight$ configuration  ($\sfrac{2}{3}$, 
3 to 4 consecutive unbalanced runs are needed, in average,  for the balancing process  to kick in.

If a particular  distribution proves to be the best for a given SCT, so far, 
the associated configuration is persisted in the KB. 
Hence, 
SCT profiles are  progressive refined,  a feature
particularly suitable to applications that operate over indiscriminate types of data-sets, and want to build on previous knowledge to progressively adjust the framework to the particularities of each of such types.

\subsubsection{The Load Balancing Process (box "\textit{Adjust workload distribution}"):}
The load balancing process serves two goals:
to refine a derived configuration, so that it better adapts to the characteristics of the computation,
and to balance the load when in the presence of fluctuations of the CPUs' load.
To achieve these goals it transfers a percentage of the workload 
from the worst to the best performing device type.
The definition of the amount of work to transfer is computed by applying a modified version of the binary search
described in Section 	\ref{sec:hw_selection}, that we name \textit{Adaptive Binary Search}.
Given the dynamic nature of the system's load distribution, 
the best solution may no longer be found in the interval under inspection.
Therefore, the adaptive binary search allows for this interval to shift sideways, so that it 
may converge to some other direction.

Morevoer, contrary to the original binary search algorithm, 
the size of the \textit{transferable partition} may also
augment in time,
in order to speed the shifting of the focal point. 
More concretely, when more then 2 shifts are performed in the same direction, the size of the 
transferable partition doubles.

\subsection{Impact on the Programming Model}
\label{sec:cpu_gpu:programming}

Multi-device execution  impacts only  the specification of the kernels' interface, more particularly on the specification 
of their parameters (instances of  \texttt{IDataType}).
%particularnamely indicate which input vectors may \emph{not} be partitioned among the devices.
%
By default, Marrow partitions  input vectors according to the locality-aware
domain decomposition 
 (Section \ref{sec:decomp}) and the set of user-defined constraints, namely the \emph{elementary partitioning unit} (defaulted to 1).
%and of the number of elements processed by each kernel thread, also defaulted to 1.
%such units processed by a single computing thread (aka OpenCL work-unit), also defaulted to 1.
%
Conversely, non-partitionable vectors must be explicitly associated to  the \texttt{COPY} data-transfer mode, 
that dispatches the enclosed data  integrally to all devices  (similarly to SkelCL \cite{skelcl_multi}).
The feature is of fundamental importance when all threads require a global snapshot of the given vector.

Complementarily, the programmer may also
 provide partition-bound values to scalar parameters,
and   specify merging functions to be applied to the partial results output by each parallel execution.
Some kernel parameters may be sensitive to the data-set's partitioning.
To deal with these cases we allow such parameters to be instantiated with special values, known to the framework's run-time system.
This feature allows the programmer to vehicle  partition-dependent information to the computation.
The supported traits are:
\begin{inparadesc}
\item[\texttt{Size}] - to instantiate the target parameter with the size of the current partition, and
\item[\texttt{Offset}]  - to  instantiate the target parameter with the offset of the partition  regarding to the entire domain.
\end{inparadesc}

The use of \emph{merging} functions is a rather common approach
to produce a single result from a set of multiple  independently computed results.
To that end, Marrow offers a set of  predefined functions (addition, subtraction, multiplication and division) and also allows for the implementation of user-defined functions.

\section{Experimental Results}
\label{sec:eval}

The purpose of this section is threefold:
1 - to quantify the performance gains yielded by the use of OpenCL fission, 
2 - to quantify the performance gains that the conjoint use of the CPU and GPU
may deliver relatively to GPU-only executions, 
and 
3 - to assess the efficiency of the work distribution and load balancing strategies proposed.

For the experiments, we resorted to five benchmarks that make use of the different skeletons available in Marrow.
The first two  make use of the \textit{Pipeline} skeleton.
\textit{Filter Pipeline} composes three image filters, namely Gaussian Noise, Solarize and Mirror. 
These  share the ability to be independently applied to distinct lines of the image to be processed. Accordingly, the image line is the 
elementary partitioning unit.
Additionally, all these kernels process two elements of the work-space  per thread.
\textit{FFT}  is a set of Fast-Fourier Transformations  adapted from the
SHOC Benchmark Suite \cite{shoc}, where FFT is  pipelined with its inversion.
The elementary partitioning unit is the size of each FFT which is 512 KBytes.
 Ergo, each device is assigned with a set of such FFTs.

The third benchmark is the  iterative \textit{N-Body} simulation supported by the \textit{Loop} skeleton.
The kernel implements the direct-sum algorithm that, for each single body computes its interaction with all the remainder. 
Therefore, there is a dependency on the whole data-set, requiring  replication to all the data-set to all devices (COPY transfer mode).
The distribution is hence performed at body level, entailing a synchronization point in-between each iteration.

The final two benchmarks are plain \textit{Map}  applications.
\textit{Saxpy}  is part of BLAS (Basic Linear Algebra Subroutines) and computes a single-precision 
multiplication of a constant with a vector added to another vector. The computation is embarrassingly parallel and does not require
any partitioning restrictions, as  each thread only operates over a single element of  each matrix. 
\textit {Segmentation}  performs a transformation over a gray-scale three dimensional image, changing its value to either white, gray or black.  Although there is no algorithmic dependencies between pixel elements, the elementary partitioning unit is set to the size of the first two dimensions so the partitioning can be performed only over the last one.
%
%
%\textit {Solarize} applies isolatedly the Solarize filter comprised in the Filter Pipeline benchmark. 

%
%The operating system is ...

\subsection{Multi-CPU  Execution}
We begin by analyzing  how our solution performs in a multi-CPU only environment. 
The objective is to assess the gains that derive from the use of the OpenCL device fission feature, for which no systematic studies can be found.
For that purpose we conducted our  experiments on a multi-processor system  comprising 64 Gbytes of RAM and
four  sixteen-core AMD Opteron 6272 @ 2.20GHz 6272 with three cache levels:  16KBytes L1 data cache per core; a unified 2MBytes L2 cache per two cores, and one unified 6MBytes L3 caches per eight cores. Four levels of  fission are supported: L1 to L3 and NUMA.

\begin{figure}
\begin{floatrow}
\capbtabbox{%
\centering
\resizebox{7cm}{!}{
\begin{tabular}{|c|c|r||r|c|r|r|}
\hline
 Benchmark & Input type  & Input argument  & Fission & Number of  &  Execution time & Execution time \\
&  & & & subdevices &    & (no fission)  \\
\hline
 & Image & 1024x1024	& 
	L3	 & 8	& 8.5	 & 9.8  \\
Filter & size & 2048x2048	& 
	L2	& 32 & 22.0 & 34.8 \\
pipeline & (pixels) & 4096x4096	& 
L2	& 32 & 65.1 & 120.3 \\
& & 8192x8192	& 
L2	& 32 & 222.8 & 377.1 \\ 

\hline
& Size & 128MB	& 
	L2 & 32 & 34.7 & 103.7 \\
FFT & of & 256MB	& 
L2 & 32 & 56.5 & 197.9 \\
& data-set & 512MB	& 
L2 & 32 & 106.4 & 423.8 \\
\hline	
& Number & 8192 &
	L2 & 32 & 35.8 & 138.4 \\
NBody & of & 16384		&
	L3 & 8 & 99.0 & 284.0 \\
 & bodies & 32768		&
	L2 & 32 & 383.4 & 1116.2 \\
& & 65536	&
	L2 & 32 & 1499.0 & 4433.6 \\
\hline	
 & Number & 1$\times 10^6$	& 
	L2 & 32 & 2.2 & 7.4 \\
 Saypy  & of & 1$\times 10^7$	&
		L2 & 32 & 23.9 & 72.1 \\
& elements & 5$\times 10^7$ &
	L2 & 32 & 102.9 & 270.8 \\
\hline	
 & Number &  1MB	& 
	L3 & 8 & 1.1 & 2.2 \\
 Segmentation  & of & 8MB	& 	
 	L3 & 8 & 4.3 & 11.8 \\
& elements & 60MB	& 
	L2 & 32 & 31.0 & 61.5 \\
\hline 
\end{tabular}
%\end{scriptsize}
}
}{
\caption{Benchmark characterization - CPU only executions}
\label{tab:benchmarkscpuonly}
}
\ffigbox{%
    \centering
    \includegraphics[width=0.9\linewidth]{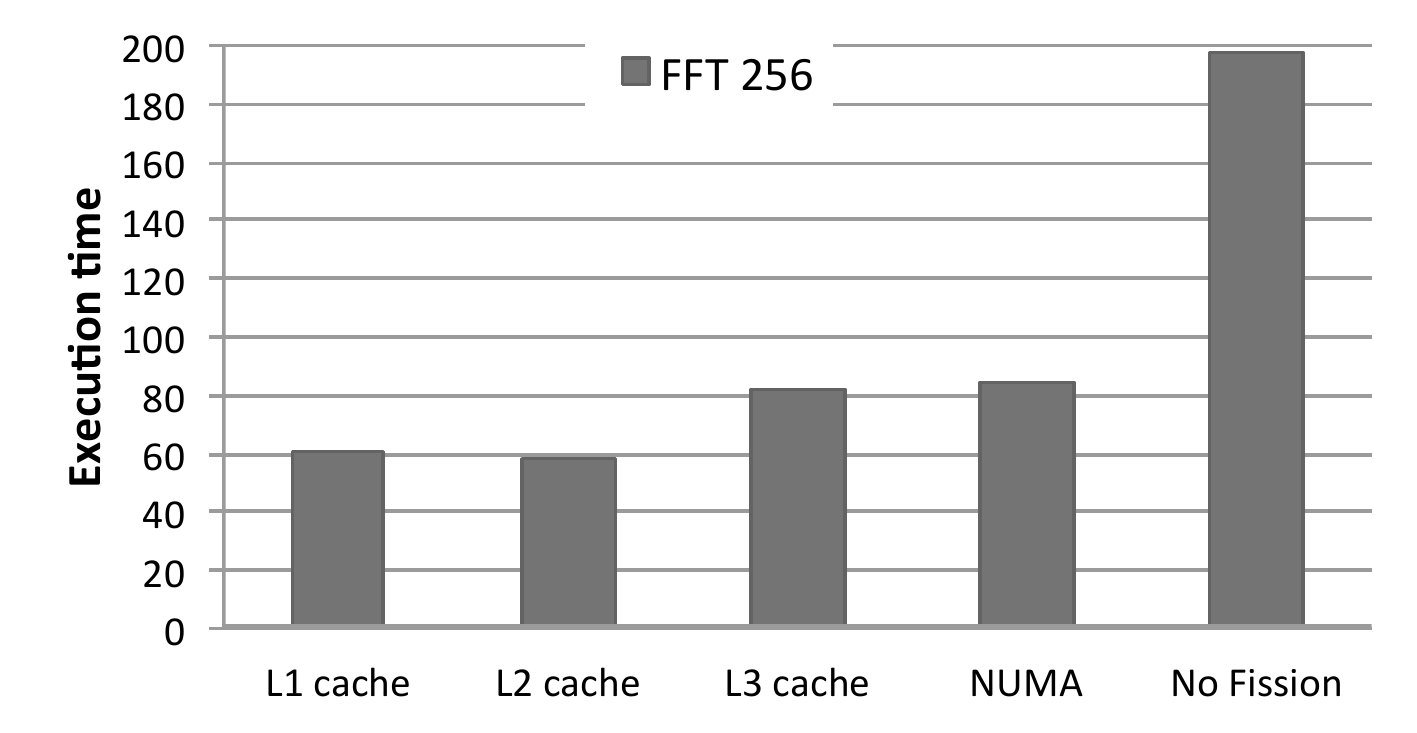}
}{
    \caption{Execution times measured in the  construction of  the profiling for FFT with 256 MB input}
    \label{fig:fft256training}
}
\end{floatrow}

\end{figure}

\begin{figure}
    \centering
    \includegraphics[width=\linewidth]{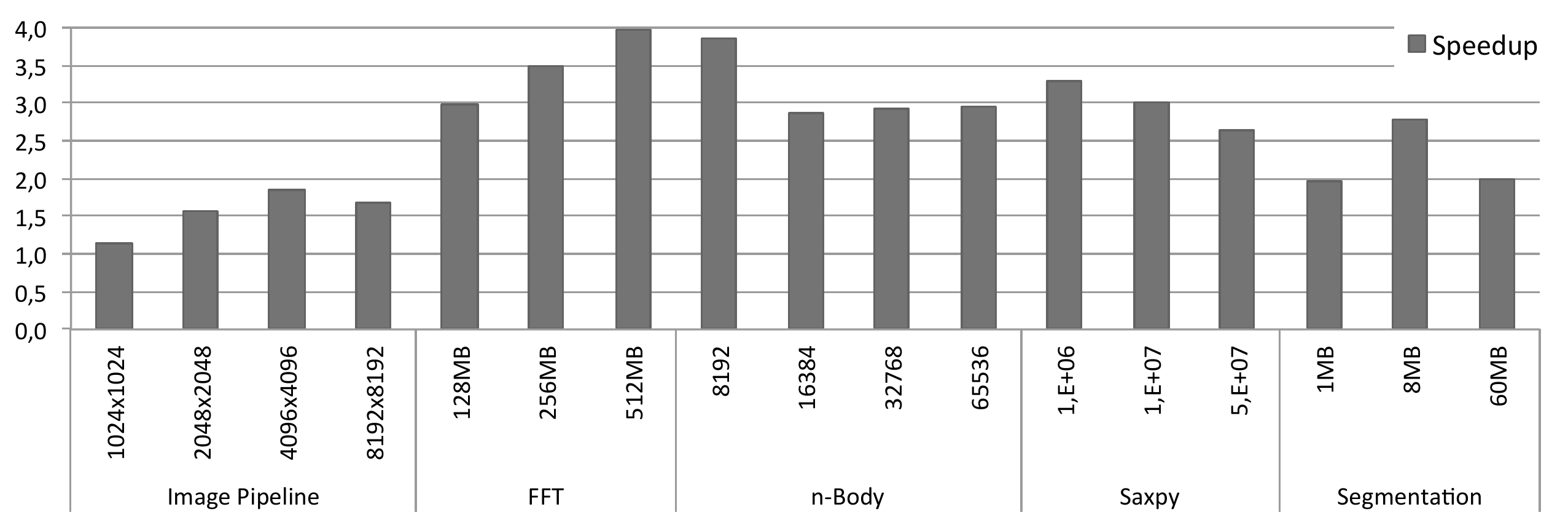}
    \caption{Speedup of Fission versus No Fission}
    \label{fig:cpu-only-all}
\end{figure}

Table \ref{tab:benchmarkscpuonly} presents the execution times  obtained for the best performing Fission configuration 
and for the original (no Fission) configuration, while Fig.  \ref{fig:cpu-only-all} depicts the speedups obtained 
by the former relatively to the latter. 
The impact is considerable, as the use of the Fission functionality provides noticeable performance gains. 
To this result  is not indifferent the fact that we are in presence of a NUMA architecture, where locality-optimizations play an important role. 
However, although the use of Fission by itself is likely to boost performance, the ideal fission level varies from benchmark to benchmark, and it is also sensitive to the size of the workload. Accordingly, the profiling process is of the foremost importance to obtain the best configuration possible.
Fig. \ref{fig:fft256training} presents the execution times measured for the multiple Fission 
configurations of the FFT 256MB benchmark.

\subsection{CPU+GPU versus GPU}

\begin{table*}
\centering
\resizebox{\textwidth}{!} {
\begin{tabular}{|c|c|r||r|c|c|c||r|c|c|c|}
\toprule
 Benchmark & Input  & Input  & \multicolumn{4}{c||}{1 GPU} & \multicolumn{4}{c|}{2 GPUs} \\
 %\cline{4-17} &   \cline{4-17} 
& type &argument &  Execution & \multicolumn{3}{c||}{Profiling Process} & Execution  & \multicolumn{3}{c|}{Profiling Process}  \\
&  && time & Configuration & Level  & Distribution   & time  & Configuration	 & Level  & Distribution  \\
&&& &  	 (fission/overlap) & of parallelism &  (GPU/CPU) &    &
(fission/overlap) & of parallelism &  (GPU/CPU)  \\
\hline
Filter & Image & 2048x2048	& 
	3.13	 & L2/4	& 10	 & 77.8/23.2 & 	
	1.96	&	L1/4	& 14 	& 79.1/20.9	\\
pipeline & size & 4096x4096	& 
	12.21	& L1/4 &	10	& 78.4/21.6 	&
	6.80	 &	L1/3	 & 12	 & 79.2/20.8  \\
& (pixels) & 8192x8192	& 
45.24	 & L1/4 &	10	& 71.4/28.6		& 
25.62	 & L1/4	& 14 	& 79.2/20.8	 \\
\hline
& Size & 128MB	& 
	21.00	& L3/4	 & 5 & 67.1/32.8	&
	12.01	& L3/4	 & 9	& 75.1/24.9	 \\
FFT & of & 256MB	& 
41.68	& L3/4	& 5 & 67.9/32.1	 	&
	23.62	 & L3/4	& 9	 & 81.6/18.4	 \\
& data-set & 512MB	& 
80.64		& L3/4	& 5 & 	68.5/31.5		& 
46.18	& L3/4	& 9 & 90.0/10.0	\\
\hline	
& Number & 16384		&
	37.17	& -/4 & 4 & 100/0 	&
	29.87	&  -/4 & 8 & 100/0   	\\
NBody & of & 32768		&
	101.56	& -/4 & 4 & 100/0  	 &
	69,63		& -/4 & 8 & 100/0   \\
& bodies & 65536	&
	356.85 		& -/4 & 4 & 100/0  &			
	200.76			& -/4 & 8 & 100/0 		 \\
\hline	
 & Number & 1$\times 10^6$	& 
	 1.83 &  L1/4	& 10 & 71.4/28.6	&
	 0.78	& L2/3 	& 12	 &85.0/15.0	  \\
 Saxpy  & of & 1$\times 10^7$	&
		16.97	&   L2/4	& 10 & 	75.3/24.7	& 
 		11.54	& L1/4	 & 14	& 87.5/12.5	\\
& elements & 1$\times 10^8$ &
	100.58  & L1/4	& 10 & 78.8/21.3	& 
	89.92  & L1/4	& 14 & 88.2/11.8 \\
\hline	
 & Number &  1MB	& 
0.47	 & L1/1 & 	6	&  90.7/	9.3&
0.41 & L3/1  &	3	& 49.5/51.5	 \\
 Segmentation  & of & 8MB	& 
 1.53		& L3/3	& 4 	& 74.2/25.7	&
 	1.17 	& L1/3	&  12	& 88.3/11.7		 \\
& elements & 60MB	& 
16.23	& L3/4	& 5 & 79.2/20.8	 &
	5.71	 & L1/3	& 12 & 87.5/12.5  \\
\hline 
\end{tabular}
}
\caption{Benchmark characterization - CPU+GPU executions}
\label{tab:benchs}
\end{table*}

For this second study the experiences were  conducted on a system  featuring 
one hyper-threaded six-core Intel(R) Core(TM) i7-3930K CPU @ 3.20GHz, comprising 6  L1 and L2 caches (one per core) and a single L3 cache, shared by all  cores;
two  AMD HD 7950 GPU devices attached to  two dedicated PCIe x16 lanes; and 32 GBytes of RAM. 
The CPU supports the L1 to L3 fission levels.

For each benchmark we have established three parameterization classes 
and two baselines, that report the time for the GPU-only accelerated execution of the benchmarks using
just one or the two available GPUs.
Table \ref{tab:benchs} presents, for each of   parameterization class, 
the baseline execution times and the results of applying 
the profiling process to both the single and dual GPU setups.

From the data depicted in this table we may observe that the best fission/overlap configuration  depends on several factors:
the actual computation, the input data-set's size and the number of devices.
Nonetheless, some patterns may be observed: 
the ideal overlap degree is directly proportional to size of the workload, 
 the ideal fission level is not highly sensible to the number of GPU devices, 
 but the level of coarse-grained parallelism is directly proportional such number, 
 and, as expected, the load assigned to the CPU  is inversely proportional to the number of GPUs.

\subsubsection{Speedup Results:}

\begin{figure}
\centering
\begin{floatrow}
\ffigbox{
\includegraphics[width=\linewidth]{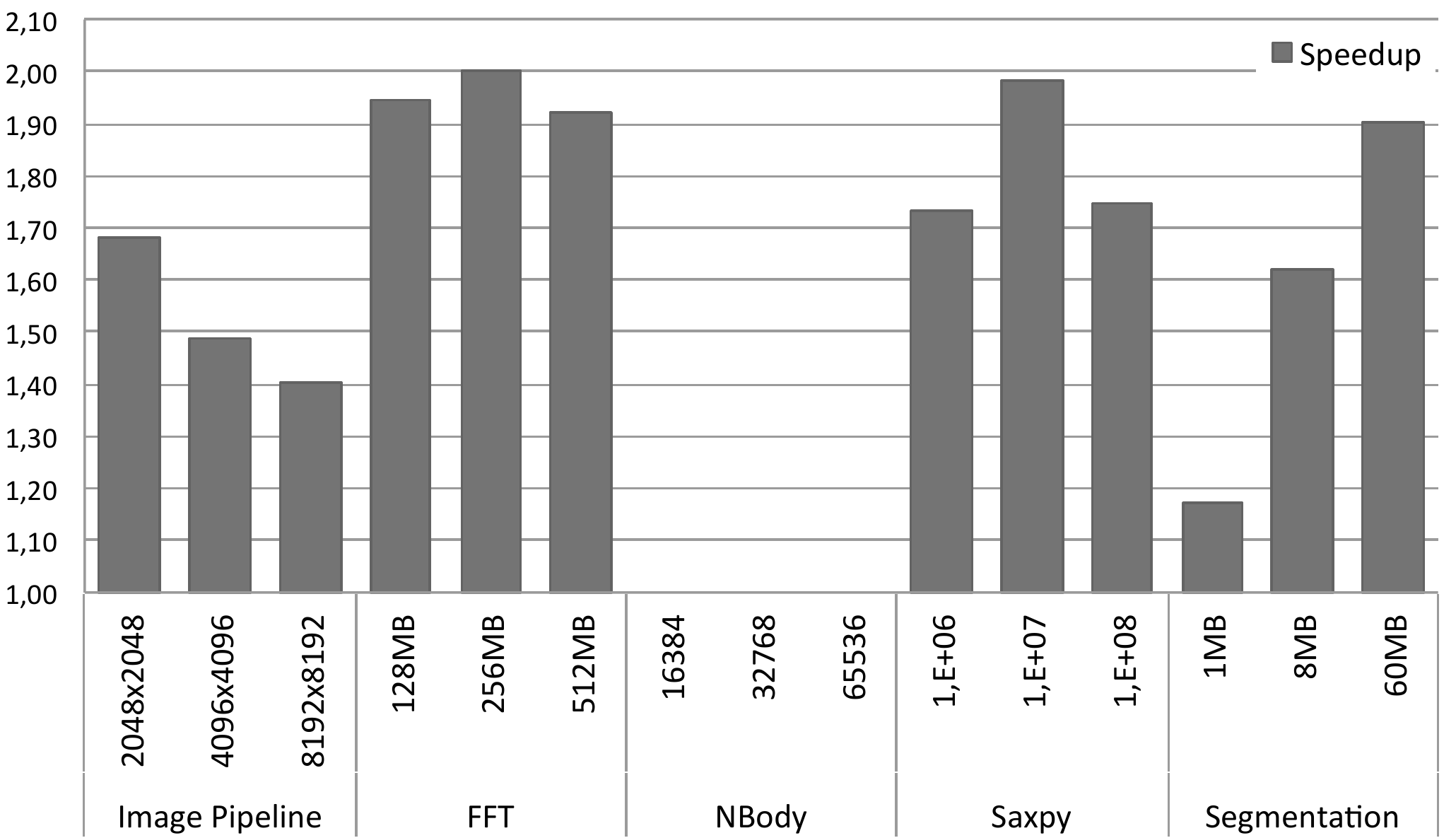}
}{
\caption{Speedup of CPU + GPU versus 1 GPU}
\label{fig:saito_1}
}
\ffigbox{
\includegraphics[width=\linewidth]{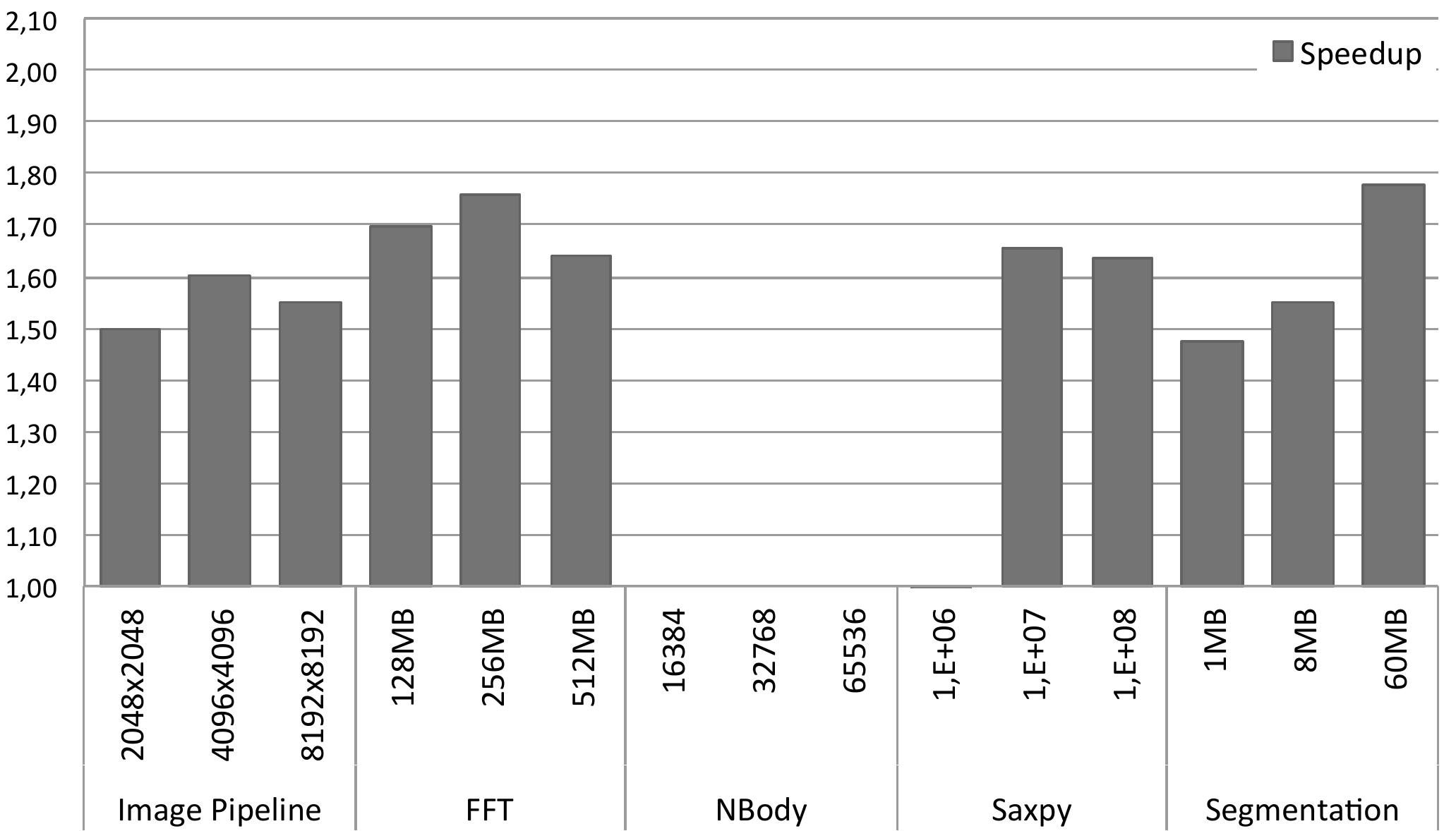}
}{
\caption{Speedup of CPU + 2 GPUs versus  2 GPUs}
\label{fig:saito_2}
}
\end{floatrow}
	\end{figure}

Figures \ref{fig:saito_1} and \ref{fig:saito_2} present the speedups obtained by the CPU/GPU ensemble when compared to the GPU-only baselines.
The results show that the use of the hybrid infrastructure  is beneficial in almost every conducted experiment; the NBody benchmark is the exception.
The impact is, naturally, more visible in the single GPU configuration, where the gains range from 111\% to 207\% (average of 172\%) , whilst in the 2 GPU
configuration they range from 100\% to 188\% (average of 156\%).

The speedups are particularly noticeable in the communication-bound computations. 
Paradigmatic examples are  Saxpy and Segmentation, where the CPU boosts the overall performance almost  twice for the single GPU scenarios. We may also observe a decay from the point where the CPU's  throughput limit is reached (higher parameterization class of Saxpy).
A similar  behavior may also be observed in the FFT benchmark.
The FFT kernels are computationally heavy but also operate upon large data-sets: 128 to 512 MBytes.
It is the ever-present trade-off between the overhead of data-transfers and the computational complexity of the computation.
The Filter Pipeline benchmark is more computation bound. Three different computations are applied over a single image transferred to the GPU.
Accordingly,  due to the effectiveness of our locality-aware domain decomposition, 
the CPU's utility is lower and mostly 
visible with smaller images, where less parallelism is required.
In the tested NBody configurations, all of the work was assigned to the GPUs, and hence 
there was no advantage on distributing work to the CPU.
The highly computation bound nature of the SCT, and the more complex execution 
model  of the Loop skeleton, that already assigns work to the CPUs and contains 
synchronization barriers, makes the use of the CPUs less relevant.

%The CPU + single GPU execution of FFT 256 

\begin{table}
\centering
\resizebox{9cm}{!}	{
\begin{scriptsize}
\begin{tabular}{|c|r|r||c|r|r|}
\hline
Benchmark &  Input parameter & $maxDev$ & Benchmark &  Input parameter & $maxDev$ \\
\hline
& 1$\times 10^6$	& 0.885 & & 2048x2048	& 0.842 \\
Saxpy	& 10$\times 10^6$	 &0.919 & Filter pipeline & 4096x4096	 & 0.855 \\
& 50$\times 10^6$	& 0.874 & & 8192x8192	& 0.846 \\
\hline
& 1MB	& 0.979 & & 128MB	& 0.846 \\
Segmentation & 8MB	& 0.863  & FFT & 256MB	& 0.825 \\
 &60MB	& 0.832 &
& 512MB	& 0.841 \\
\hline
\end{tabular}
\end{scriptsize}
}
\caption{Maximum deviation}
\label{tab:maxdev}
\end{table}

\subsubsection{Efficiency of the Work Distribution and Load Balancing Strategies:} 
		
Our initial study aims to assess if there is a maximum deviation threshold
that can be generally applied, so that framework may operate without resorting to the load balancing 
process under stable load conditions.
We selected a subset of studied benchmark applications, in order to 
determine a value for  $maxDev$ that allows for 500 executions of each 
benchmark to execute without triggering the load balancing process.
To guarantee the stability premises, the framework is the sole user application running in the system.
The  results 	 in Table \ref{tab:maxdev} allow us to conclude that, in general, [0.8, 0.85] is an adequate range of 	values for $maxDev$.
These guarantee that all concurrent executions are within 80\% to 85\% 
of  the best performing one.
The framework will only enter the load balancing mode
when  it recurrently identifies  executions,  that do not meet this requirement,  in a small number  of iterations.

Next, we selected the Filter Pipeline  benchmark to evaluate how does our configuration derivation behaves
in the presence of different input data-sets.
We begin with an empty KB, and populate it as  the benchmark is successively applied to images \textit{Image 0} to \textit{Image 7}, 
Thus, when \textit{Image $i$} is executed the KB contains knowledge about images \textit{Image 0} to 
\textit{Image $i-1$}.

\begin{figure}
\begin{floatrow}
\ffigbox{
\includegraphics[width=\linewidth, height=4cm]{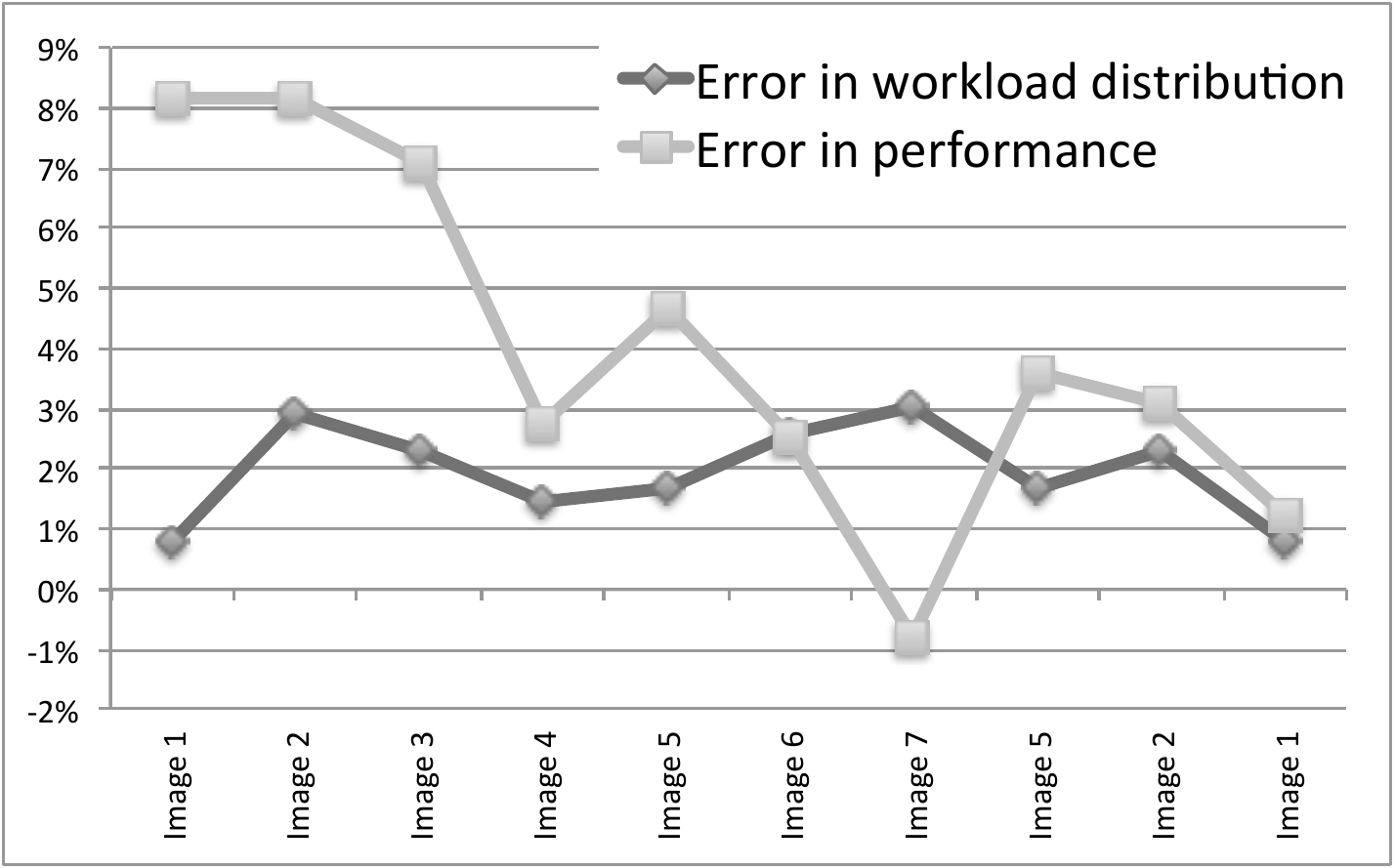}
}{
\caption{Evolution of the error (in \%) in the application of Filter Pipeline to images 1 to 7}
\label{fig:image_pipe:error}
}
\ffigbox{
\includegraphics[width=\linewidth, height=4cm]{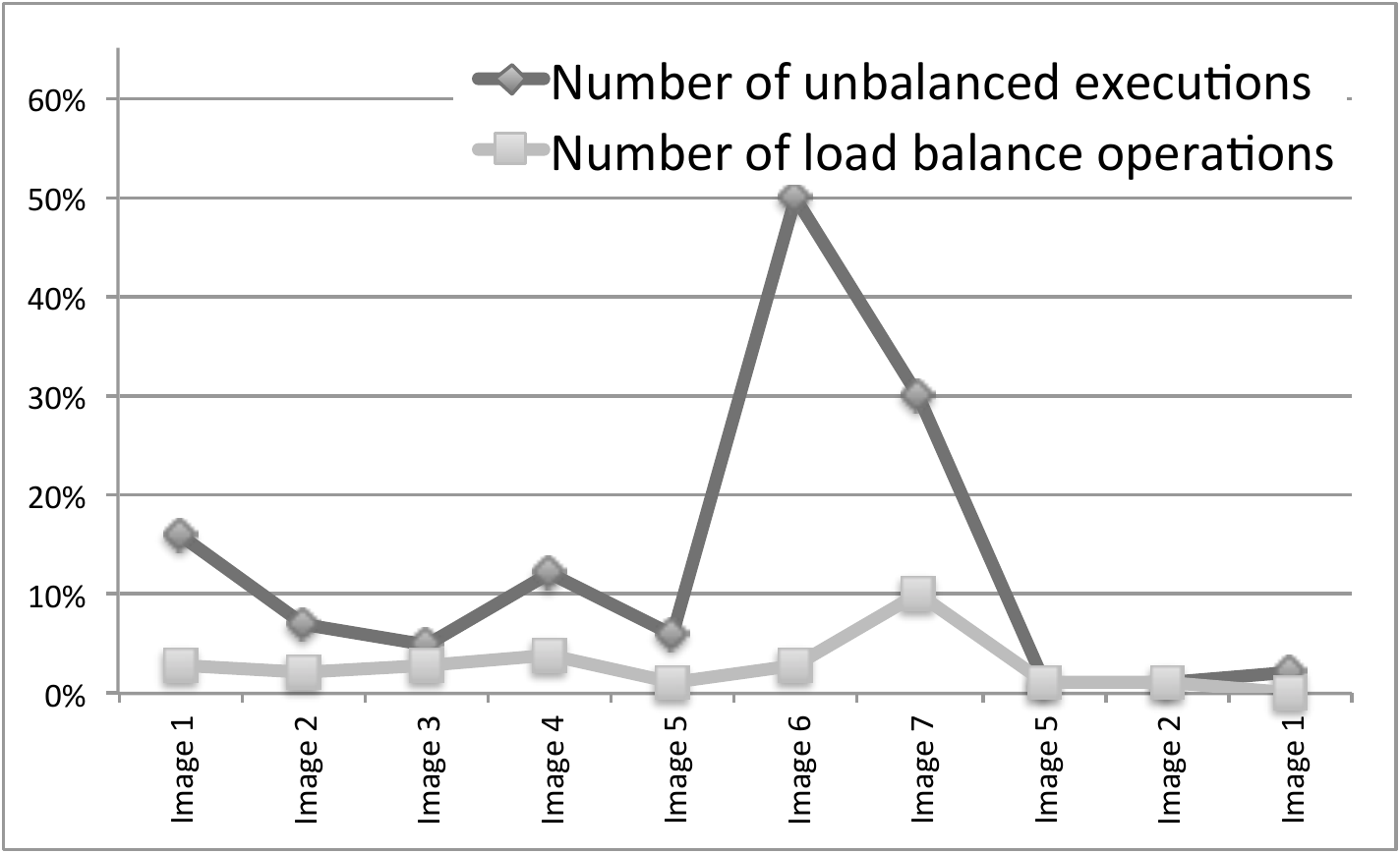}
}{
\caption{Impact of the load balancing process in the application of Filter Pipeline to images 1 to 7}
\label{fig:image_pipe:lb}
}
\end{floatrow}
\end{figure}

To establish individual baselines, we independently ran the profile building process  
 for each image. The left-hand side of Table \ref{tab:pipeline} presents such results.
Then, beginning with a  KB containing only the profile of \textit{Image 0}, 
and with the profile construction option switched off, we 
 successively applied the benchmark to images \textit{Image 1} to \textit{Image 7}. 
The benchmark was configured to run 100 times, with 
$maxDev$ set at 0.85, so that
 we could also evaluate the impact of the load balancing process.
The right-hand side of Table \ref{tab:pipeline} depicts the configuration derived from the KB, 
the number of unbalanced executions (deviation $>$ 0.85), the number of times the load balancing process was triggered,
the obtained distribution (which is persisted in the KB), and the performance yield by this final configuration.
Additionally, Fig. \ref{fig:image_pipe:error} illustrates
the evolution of the error in the workload distribution and performance of the derived configuration relatively to the
one obtained via the profile construction process, and
Fig.  \ref{fig:image_pipe:lb} illustrates
the evolution in the number of unbalanced executions and in the number of times the load balancing process in triggered.
With the purpose of evaluating 
the steadiness of the derived results, we apply the benchmark a second time
to images  \textit{Image 5},  \textit{Image 2}  and  \textit{Image 1}.

\begin{table}
\centering
\resizebox{\textwidth}{!} {
\begin{tabular}{|c|r||r|r||r|r|r|r|r|}
\hline
Image & Image 	& \multicolumn{2}{c||}{Profile Construction} &  \multicolumn{5}{c|}{Profile Derivation} \\
id & size & 	 & & {Derived distribution}  & Unbalanced & Load balance & {Persisted distribution} & Execution 	 \\
 &  & 	 GPU (\%)	 & Execution time &   GPU (\%)	&  executions & operations & 	  GPU (\%)	&  time \\
\hline
Image 0	& 1024x1024	& 87.5\% &  1.10  & & & & &  \\				
Image 1 & 	4288x2848	 &  92.8\%  & 10.68 & 87.5\% & 16 &  3 & 93.5\% & 11.55 \\
Image 2& 	512x512	& 94.8\% & 0.53 & 93.5\%	& 7 & 2 & 92.0\% & 0.61 \\
Image 3 & 	8192x8192	& 93.8\% &   45.24 & 94.5\%	& 5 & 3 & 91.6\%		& 48.45 \\  	
Image 4& 	1800x1125	&  99.9\% & 1.90 & 91.7\%	& 12 & 4 &  98.4\%	& 1.95 \\  
Image 5	& 2048x2048	&   92.7\% & 3.13 & 92.4\%	& 6 & 1 & 94.3\%	& 3.28 \\ 
Image 6	& 256x512	 &  89.6\%  &  0.43 &  90.4\%	& 50 & 3 & 91.9\% & 0.44 \\ 	
Image 7	 & 1440x900	&  94.9\% & 1.35 & 92.7\% & 30 & 10 & 92.0\% &	1.33 \\
\hline
\end{tabular}
}
\caption{Profile construction versus profile derivation}
\label{tab:pipeline}
\end{table}

What may be initially concluded is that we are able to derive suitable configurations for the different images, even the training set
is not very extensive. With very few load balancing operations we are able to keep the performance error below 5\%
after the first three images. To that result contributes the fact that the error associated to the derived workload distribution is always kept under 3\%.
Regarding the impact of the load balancing process, it is usually triggered less than 4 times in 100.
Exceptions arise in small images when the initial derivation is not spot on.
Such behaviour can be observed in \textit{Image 7}, whose executions are shift between balanced and unbalanced 
(relatively to the  \textit{maxDev} parameter),
triggering the load balancer 10 times until a steady configuration is found.

For our last experiment, we selected the 128  parameterization class of the FFT benchmark to study how the framework adapts to load fluctuations in the CPU.
From Table \ref{tab:benchs} we have that the initial workload distribution is GPU $\leftarrow$ 75.5\%	 and CPU $\leftarrow$ 24.5\%.

\begin{figure}
\includegraphics[width=\linewidth]{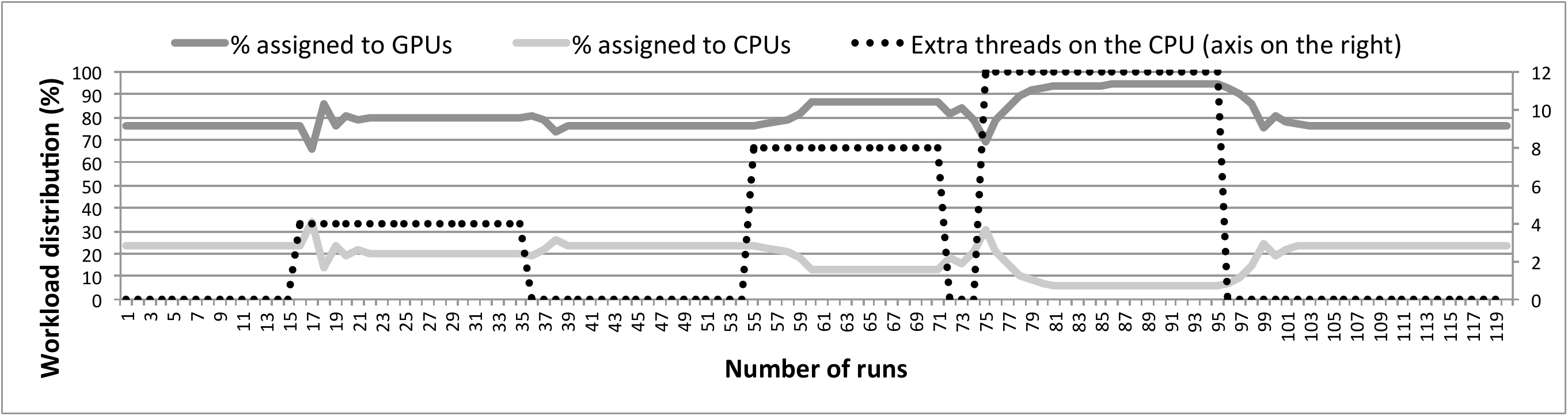}
\caption{FFT 128 benchmark subjected to CPU load fluctuations}
\label{fig:load}
\end{figure}

To introduce the load fluctuation on the CPU we implemented an application  that spawns 
a given number of
software threads,
each running a computationally heavy algebraic problem. 
Fig. \ref{fig:load} depicts the framework's adaptation to the 
sudden increase in the CPU's load.
Once the unbalancing has been detected, the load balancing process increasingly assigns more work 
to the best performing device type (the shifting phase) until it finds the interval where the standard binary search can be smoothly applied.
The shifting phase is usually abrupt but quick - 1 to 4 runs -  
while the in-depth binary search draws a smoother line and may take a little more than 10 runs.

\section{Related Work}
\label{sec:related}

SkePU \cite{skepuadaptative}, SkelCL \cite{skelcl_multi}, Muesli \cite{muesli} Bolt \cite{bolt}, and Thrust \cite{thrust} are   skeleton/template frameworks for GPU computing. Their focus is on data-parallel skeletons (map and reduce are the more common). None of them support skeleton nesting, thus no compound behaviors can be offloaded to the GPU.
Heterogeneity support in these frameworks comes in two flavors. The first approach is to include, at language level, constructions to determine where the computation must take place. This is the case of Muesli where 
only a subset of the  available data-parallel skeletons may be used to offload computations to the GPU. 
The second approach obliges the programmer to direct the compilation at either CPUs or GPUs. This category includes SkePU and Thrust, which feature multiple back-ends, for either GPGPU (CUDA and/or OpenCL) or shared-memory parallel programming (OpenMP and/or Intel TBB). These are however mutually exclusive and must be selected at compile-time. 
The SkePU library can be combined with StarPU \cite{StarPU}, a heterogeneous computing scheduler. Although StarPU has the capability to schedule tasks to run on multi-core CPUs and GPUs simultaneously, when a task is submitted to SkePU, the best performing device for the given input size is selected, but only one device will execute the job. 
The work presented in \cite{adaptativegpu}  also considers different devices without the programmer's intervention, selecting the best device to run the computation but never resorting to  CPU/GPU
wide computations.

StreamIt \cite{streamit-gpu} and Lime \cite{lime} provide linguistics constructions to express task and data-parallel computations. 
Lime is an  object-oriented language  almost completely compatible with Java. Among the  features that it adds to the Java  syntax are operators for pipelining, map and reductions. The combination of pipes with the data-parallel operations are automatically identified, and the subsequent computations offloaded to the GPU. However, a number of restrictions are imposed upon the code that can be offloaded.
In turn, StreamIt is a stream processing language based on  three main computation structuring constructs: pipeline, split-join and feedback-loop - with which disciplined graphs may be built.
Recently, it has been equipped with a CUDA back-end. A static analysis performed at compile-time transforms the graph so it can be distributed among multiple GPUs.
A similar approach is followed by Harmony \cite{harmony}, which generates a set of kernels from a higher-level representation by building a dependency graph. The resulting kernels may be scheduled onto different types of processing units (CPUs, GPUs and FPGAs).

Dandelion \cite{dandelion} offers  LINQ-based programming model  for  CPU/GPU systems, 
extended with other constructions, such as loops.
A specialized compiler generates  code for both CPU and GPU  from the source code,
while the run-time system transparently offloads computations to the GPU, when
the workload so justifies.

In \cite{mapreducecpugpu}, 
the authors propose a runtime infrastructure for  MapReduce computing on  CPU/GPU systems.
Two different work distribution methods are presented, 
one that  schedules map tasks across the processing units (PUs), and a second that places the map and the reduction stages on  different PUs,
much like what we allow in our MapReduce skeleton.
The dynamic work distribution tries to adjust the task block sizes  
for subsequent executions of  tasks in the same application.

Qilin \cite{qilin} is a heterogeneous 	programming system
that generates code for both CPUs and GPUs.
It follows an approach close to ours  
for a single kernel in single CPU/GPU nodes.
It builds a profile of the kernel and performs an adaptive mapping of the computation onto the target hardware.
The work analyzes the sensitivity of the mapping to hadware and software changes, but not to changes in the system's load.
SKMD \cite{LeeSPM13} also  dynamically distributes  a  single data-parallel kernel across multiple asymmetric CPUs and GPUs.
The focus is on optimizing data transfer cost and determining ideal 
work-group sizes, having in mind the performance variance of the multiple devices.

Finally, in \cite{tecnicoppam}, the presented solution allows for work-partitioning among devices based on a performance model updated at run-time. This solution, however, requires the existence of multiple versions of the program, one for 
each different device present in the system.

\section{Concluding Remarks}
\label{sec:conclusions}

We presented a systematic approach for the cooperative
multi-CPU/multi-GPU execution of  Marrow computations.
We  proposed a locality-aware domain decomposition of Marrow skeleton trees
that promotes data-locality but, simultaneously,  allows for the multiple OpenCL kernels to 
be executed under different work-group  configurations, as long as  communication compatibility is assured.
We  also proposed profile-based workload distribution and load balancing strategies, that 
build from past runs of a given computational tree to derive suitable configurations
for subsequent executions.
The experimental results show that our approach brings significant speedups  over GPU-only executions.
Moreover, they  also provide some insight on the framework's ability to adapt to  data-sets of different
sizes and to fluctuations on the CPU's load.

Regarding future work, our focus is on the  improvement of our configuration derivation
via a static analysis of the kernels' code.
We are also working on the incorporation of other types of accelerators
and on the combination of OpenCL with other technologies, such as OpenMP.

\bibliographystyle{elsarticle-num}
\bibliography{main}

\end{document}